\documentstyle[prb,aps,preprint,epsf]{revtex}
\tightenlines

\begin{document}
\draft

\title{
Magnetoresistance Effect in Spin-Polarized Junctions of Ferromagnetically Contacting Multiple Conductive Paths: Applications to Atomic Wires and Carbon Nanotubes }
\author{
Satoshi~\textsc{Kokado}$^1$ 
\footnote{Electronic mail:satoshi-kokado@aist.go.jp}
and 
Kikuo~\textsc{Harigaya}$^{1,2}$
\footnote{Electronic mail:k.harigaya@aist.go.jp}
}
\address{
$^1$Nanotechnology Research Institute, AIST, Tsukuba 305-8568, Japan \\
$^2$Synthetic Nano-Function Materials Project, AIST, Tsukuba 305-8568, Japan
}
\date{\today}

\maketitle

\begin{abstract}
For spin-polarized junctions 
of ferromagnetically contacting multiple conductive paths, 
such as ferromagnet (FM)/atomic wires/FM 
and FM/carbon nanotubes/FM junctions, 
we theoretically investigate spin-dependent transport 
to elucidate the intrinsic relation between 
the number of paths and conduction, 
and to enhance the magnetoresistance (MR) ratio. 
When many paths are randomly located between the two FMs, 
electronic wave interference between the FMs appears, 
and then the MR ratio increases with increasing number of paths. 
Furthermore, at each number of paths, 
the MR ratio for carbon nanotubes becomes larger 
than that for atomic wires, 
reflecting 
the characteristic shape of points in contact with the FM.
\end{abstract}

\narrowtext
\newpage

Recently, 
the magnetoresistance (MR) effect in spin-polarized junctions of 
a ferromagnetically contacting single conductive path, 
which is ferromagnet (FM)/path/FM junctions 
like Co/carbon nanotube/Co junctions~\cite{Tsukagoshi,Zhao1,Zhao2,th1}, 
has been reported. 
When the MR ratio was defined as 
$(\Gamma^{\mbox{\tiny (P)}} - \Gamma^{\mbox{\tiny (AP)}})/
\Gamma^{\mbox{\tiny (P)}}$, 
with $\Gamma^{\mbox{\tiny (P)}}$ and $\Gamma^{\mbox{\tiny (AP)}}$ 
being the conductances of parallel (P) 
and antiparallel (AP) magnetization configurations of the FMs, 
its magnitude for Co/carbon nanotube/Co 
was experimentally observed to be 9\%~\cite{Tsukagoshi}, 
23\%~\cite{Zhao1}, and 26\%~\cite{Zhao2} at 4.2 K. 
In the theoretical study 
based on Green's function method, 
the maximum value of the MR ratio was evaluated to be 20\%~\cite{th1}. 
From the viewpoint of spintronics, however, 
larger MR ratios than those values are strongly desired, 
because 
conventional systems with film spacer, 
FM/Al-O/FM junctions~\cite{conven1,conven2}, 
have much larger MR ratios 
in spite of using a FM 
with almost the same spin polarization rate~\cite{SP2}.

The difference in the MR ratio between 
the film and the single path systems 
originates from 
that in the number of conductive paths 
connecting the two FMs or that of conductive mechanisms. 
The former has many conductive paths 
and tends to exhibit coherent conduction (CC)~\cite{CC,semi}, 
in which the intralayer momentum of the FMs 
is conserved in the transmission between them, 
while the latter exhibits incoherent conduction (IC)~\cite{IC}, 
where the momentum is not conserved.

Our interests are 
in how the number of paths influences 
the coherence of conduction  
and 
how the coherence contributes to the enhancement of the MR ratio. 
Microscopic theoretical studies 
on such questions 
have not been reported so far. 
By introducing many paths in the path systems, 
we may find an intrinsic relation between the paths 
and the coherence of conduction. 
Furthermore, 
if the MR ratio in the path systems is successfully enhanced by many paths, 
our study will significantly contribute to 
the development of future nanowire or nanotube MR devices.

In this work, 
we investigate spin-dependent transport properties 
of FM/many paths/FM junctions, 
where many conductive paths are randomly located 
between the two FMs. 
Using Green's function technique~\cite{Todo}, 
we obtain an expression of conductance 
having not only the IC term but also the CC term, 
which appears as a result of 
electronic wave interference between the two FMs. 
In applications to atomic wire and carbon nanotube systems, 
we find that the MR ratios increase 
with increasing number of paths because of an increase of CC. 
We will demonstrate that 
such electronic wave interference brings about 
the enhancement of MR ratios in path systems.

First, let us derive an expression of the conductance 
of FM/many paths/FM junctions shown in the left panel of Fig. 1. 
Here, the FM is an electrode, 
the path can be, for example, an atomic wire or a nanotube, 
and many paths are randomly located 
under the assumption that 
contact points between their edges and the FM 
exist parallel to the $x$-axis and at $y$=0, 
and current flows in the $z$ direction. 
Keeping the qualitative discussion in mind, 
we assume that 
all parts consist of atoms with a single orbital, 
and the FM has a simple cubic structure with 100 monolayers. 
Furthermore, 
the intralayer direction ($xy$-plane) of the FM 
is regarded as an infinite system 
and is set to have the periodic boundary condition. 
The system is described using a tight-binding model 
with the nearest neighbor transfer integral.

Within Green's function technique~\cite{Todo}, 
the conductance for spin $\sigma$ at zero temperature 
is written as 
\begin{eqnarray}
&&\Gamma_\sigma = 
\frac{2\pi e^2}{\hbar}
{\rm Tr} [D_{L,\sigma} T_\sigma^\dag D_{R,\sigma} T_\sigma], 
\end{eqnarray}
with 
$T_\sigma =V_\sigma + V_\sigma G_\sigma^\dag V_\sigma$, 
$G_\sigma = (E_{\mbox{\tiny F}} - H_{\sigma} + {\rm i} 0^+)^{-1}$, 
$D_{j,\sigma} = \frac{-1}{\pi}{\rm Im}
\left( E_{\mbox{\tiny F}} -  H_{j,\sigma} + {\rm i} 0^+ \right)^{-1}$, 
$ H_{\sigma}= \sum_l  H_{\ell,\sigma} + V_\sigma$, 
where $ H_{\sigma}$ is the Hamiltonian for the whole systems, 
$ H_{\ell,\sigma}$ is the Hamiltonian for an $\ell$ part with 
$\ell$ being a suffix for the left FM, paths, or the right FM, 
$V_\sigma$ is coupling between the FM and the paths, 
$D_{j,\sigma}$ ($j$=L or R) is the density of states (DOS) operator, 
and $E_{\mbox{\tiny F}}$ is the Fermi level 
at which conduction electrons exist. 
The unit of $\frac{2 \pi e^2}{\hbar}$=1 is adopted below.

When the atomic orbital at the contact points between the FM and the paths 
is represented as $|x,y,z \rangle$ indexed by $x,y,z$ coordinates, 
the conductance is rewritten as 
\begin{eqnarray}
&&\Gamma_\sigma = 
\sum_{x,x',x'',x'''} 
\langle x,0,L|D_{L,\sigma}|x',0,L \rangle 
\langle x',0,L|T_\sigma^\dag|x'',0,R \rangle \nonumber \\
&&\hspace*{1.cm} \times \langle x'',0,R|D_{R,\sigma}|x''',0,R \rangle 
\langle x''',0,R|T_\sigma|x,0,L \rangle, 
\end{eqnarray}
where $L~(R)$ is the $z$ coordinate 
at the interfacial layer in the left (right) FM. 
Transference of state represents the circular propagation, 
left FM $\to$ right FM $\to$ left FM, 
and $T_\sigma$ connects the two FMs. 
We here use an approximation in which 
$T_\sigma$ is finite only for the same $x$ coordinate 
between the left and right FMs, 
and it is independent of $x$. 
The validity of the approximation will be described in applications. 
Using the Bloch wave at the interfacial layer in the FM, 
$| {\bf k},L(R) \rangle$, where ${\bf k}$ $[=(k_x,k_y)]$ 
is the intralayer wave vector of the FM, we obtain 
\begin{eqnarray}
&&\Gamma_\sigma = \frac{1}{(N_x N_y)^2} |T_\sigma |^2  \nonumber \\
&&\hspace*{0.8cm}
\times \sum_{x,x'} \sum_{{\bf k},{\bf k}'}
{\rm e}^{-{\rm i} (k_x - k_x') (x - x')} 
D_{L,\sigma} ({\bf k}) D_{R,\sigma} ({\bf k}'), 
\end{eqnarray}
with 
$D_{L(R),\sigma} ({\bf k})=
\langle {\bf k},L(R)|D_{L(R),\sigma}|{\bf k},L(R) \rangle$, 
where 
$|T_\sigma|^2$ denotes the transmission coefficient 
including information on the paths, 
$N_{x(y)}$ is the number of atoms on the $x~(y)$-axis 
at the interfacial layer in the FM, 
and $(k_x - k_x') (x - x')$ corresponds to 
difference in the phase of the Bloch wave between the left and right FMs.

Now, based on randomly located paths, 
we average $\Gamma_\sigma$ over the distribution of the paths, 
that is, $\left\langle \Gamma_\sigma \right\rangle_{\mbox{\tiny path}}$. 
We utilize 
$\langle 
\sum_{x,x'} {\rm e}^{-{\rm i} (k_x - k_x') (x - x')} 
\rangle_{\mbox{\tiny path}}
= 
\langle \sum_{x,x'} \delta_{x,x'} + 
\sum_{x \ne x'} {\rm e}^{-{\rm i} (k_x - k_x') (x - x')} 
\rangle_{\mbox{\tiny path}}
= 
N_{\mbox{\tiny path}} + 
N_{\mbox{\tiny path}}(N_{\mbox{\tiny path}} -1 ) \delta_{k_x,k_x'}$
~\cite{second},
where $N_{\mbox{\tiny path}}$ is the number of paths. 
The first term means that 
the difference in the phase disappears 
when the electron uses the same path in the circular propagation. 
The second term represents 
the propagation with different paths. 
Correlation of the Bloch wave between the left and right FMs 
becomes large for $k_x = k_x'$, 
while it is negligibly small for $k_x \ne k_x'$. 
It is interpreted just as the result of the interference 
between the electronic wave in the left FM and that in the right FM. 
The conductance per $N_{\mbox{\tiny x}}$ becomes 
\begin{eqnarray}
&&\frac{\Gamma_\sigma }{N_{\mbox{\tiny x}}} =
\frac{N_{\mbox{\tiny path}}}{N_{\mbox{\tiny x}}}
|T_\sigma|^2 D_{L,\sigma}D_{R,\sigma}  \nonumber \\
&&\hspace*{1cm}+  \frac{
N_{\mbox{\tiny path}}^2-N_{\mbox{\tiny path}}}{{N_x}^2} 
|T_\sigma|^2 
\frac{1}{N_{\mbox{\tiny x}}}
\sum_{k_x} D_{L,\sigma}(k_x)D_{R,\sigma}(k_x), 
\end{eqnarray}
with 
$D_{L(R),\sigma}= \frac{1}{N_x N_y}\sum_{{\bf k}}D_{L(R),\sigma}({\bf k})$ 
and $D_{L(R),\sigma}(k_x) =\frac{1}{N_y} \sum_{k_y}D_{L(R),\sigma}({\bf k})$. 
The first and second terms represent 
IC with nonconservation of ${\bf k}$ 
and CC with conservation of $k_x$, respectively. 
For $N_{\mbox{\tiny path}}$=1, 
only the IC term remains, and it results in an expression 
obtained straightforwardly for the single path systems~\cite{Todo}. 
With increasing $N_{\mbox{\tiny path}}$, 
the CC term increases rapidly, 
meaning that 
the interference effect is enhanced by 
an increase of the propagation between the two FMs. 
Also, using this expression, we calculate the MR ratio, 
which is defined as $(\Gamma_\uparrow^{\mbox{\tiny (P)}} 
                   +  \Gamma_\downarrow^{\mbox{\tiny (P)}} 
                   -  \Gamma_\uparrow^{\mbox{\tiny (AP)}} 
                   -  \Gamma_\downarrow^{\mbox{\tiny (AP)}}) 
                   /( \Gamma_\uparrow^{\mbox{\tiny (P)}} 
                   +  \Gamma_\downarrow^{\mbox{\tiny (P)}})$.

We simply explain the MR ratios for the IC and CC terms 
by paying attention to only the DOSs (see right panel of Fig. 1). 
In the case of IC, 
based on the conductance with the product of DOSs, 
the MR ratio can be estimated using the Julliere model~\cite{jul}. 
The MR ratio for the Co electrode 
becomes merely about 21\%~\cite{Tsukagoshi}. 
On the other hand, in the case of CC, 
the conductance is related to $k_x$-dependent DOS, 
as shown by the shaded region in this Fig. 1. 
When DOS with $k_{x,\uparrow}^*$ and $k_{x,\downarrow}^*$ 
at the left FM is in the vicinity of $E_{\mbox{\tiny F}}$, 
the DOS at the right FM is 
in the vicinity of $E_{\mbox{\tiny F}}$ in the P case 
and absent in the AP case. 
Then, the conductance with $k_{x,\uparrow}^*$ and $k_{x,\downarrow}^*$ 
is finite in the P case, 
while it is almost zero in the AP case. 
This obvious difference in the conductance between the P and AP cases 
leads to a large MR ratio.

As an application to realistic systems, 
we first consider FM/atomic wires/FM junctions, 
where the atomic wire consists of $n$ atoms, and 
the $z$ coordinate of the left (right) edge of the atomic wire 
is indexed by 1 ($n$). 
We here obtain an exact expression of $T_\sigma$, 
which finally becomes 
$\langle x,0,L| T_\sigma |x',0,R \rangle = 
\langle x,0,L| V_\sigma G_\sigma V_\sigma |x',0,R \rangle =
v^2 \langle x,0,1 | G_\sigma |x',0,n \rangle$ 
with $v$ being a transfer integral between the FM and the atomic wire, 
by solving the Dyson equation 
$G_\sigma=g_\sigma + g_\sigma V_\sigma G_\sigma$, 
with 
$G_\sigma=( E_{\mbox{\tiny F}} - H_{\sigma} + {\rm i} 0^+)^{-1}$ 
and $g_\sigma=( E_{\mbox{\tiny F}} - 
\sum_\ell H_{\ell, \sigma} + {\rm i} 0^+)^{-1}$, 
i.e., 
\begin{eqnarray}
&&\tilde{T}_\sigma = 
v^2 g_{1n,\sigma} (\tilde{1} + \tilde{\Pi}_\sigma )^{-1}, \\
&&\tilde{\Pi}_\sigma = 
- v^2 (g_{11,\sigma} \tilde{g}_{L,\sigma} 
+ g_{nn,\sigma} \tilde{g}_{R,\sigma} ) \nonumber \\
&&\hspace*{0.9cm}+v^4 
(g_{11,\sigma} g_{nn,\sigma} \tilde{g}_{R,\sigma} \tilde{g}_{L,\sigma}  
- g_{1n,\sigma} g_{n1,\sigma} \tilde{g}_{R,\sigma} \tilde{g}_{L,\sigma} ), 
\end{eqnarray}
with 
$g_{ij,\sigma}=\langle x,0,i | g_\sigma | x,0,j \rangle$ for $i,j$=1 or $n$, 
$\tilde{g}_{L(R),\sigma}= \sum_{x,x'} | x \rangle \frac{1}{N_x N_y} 
\sum_{\bf k} {\rm e}^{-{\rm i} k_x (x - x')} 
g_{L(R),\sigma} ({\bf k}) \langle x' |$, 
$g_{L(R),\sigma} ({\bf k})=
\langle {\bf k},L(R) | g_{\sigma} | {\bf k},L(R)  \rangle$, 
and $\tilde{1}=  \sum_{x} | x \rangle \langle x|$, 
where 
$|x \rangle$ is introduced for a matrix representation 
about the $x$ coordinate. 
Here, 
$\tilde{T}_\sigma$ is a matrix including $\tilde{\Pi}_\sigma$ 
which has off-diagonal elements between different paths, 
reflecting the propagation throughout all paths. 
For $\tilde{g}_{L(R),\sigma}$, 
we now take into account only the diagonal elements, 
because those elements are much larger than the off-diagonal ones 
owing to the disappearance of the phase factor, $k_x (x - x')$. 
Therefore, $\tilde{T}_\sigma$ results in a diagonal matrix 
having an expression for the single path systems as the matrix element, 
which is independent of $x$. 
It should also be mentioned that 
$v$ is assumed to be smaller than 
the transfer integrals in the FM and the atomic wire parts 
for reasons such as differences of types between two orbitals 
and imperfect lattices matching at the interface. 
Therefore, the contribution of $\tilde{\Pi}_\sigma$ is small.

We use the following parameters. 
The number of atoms in the single wire is 20, 
the transfer integrals in the FM and the wire parts are $t$ ($t<0$), 
and the transfer integral between the FM and the wire 
is $v=0.1t$. 
When $E_{\mbox{\tiny F}}$=0, 
the on-site energy of up (down) spin for the FM 
is 3.8$|t|$ (4.2$|t|$)~\cite{ex}, 
and that of the wire is 0, 
which corresponds to the conductive wire.

Figure 2 shows the MR ratio vs $\frac{N_{\mbox{\tiny path}}}{N_x}$. 
The MR ratios for $N_{\mbox{\tiny path}} > 1$ 
are larger than that for $N_{\mbox{\tiny path}} = 1$, 
and it gradually increases with increasing $N_{\mbox{\tiny path}}$. 
This behavior originates from the contribution of the CC term. 
As the inset shows, 
the CC term causes a large difference between the P and AP cases, 
while the IC term exhibits little difference between them. 
Furthermore, 
such a CC term and its difference between the P and AP cases increase 
with increasing $N_{\mbox{\tiny path}}$. 

As the second application, 
we consider single-walled 
($N_{\mbox{\tiny cir}}$, 0) zigzag carbon nanotubes~\cite{cnt}, 
where $N_{\mbox{\tiny cir}}$ is the number of unit cells in the zigzag edge. 
Each edge carbon atom of the nanotube is assumed to
interact with its nearest atom in the cubic lattice of the FM. 
The interaction is denoted as the transfer integral $v$. 
In a similar way to that in the case of the atomic wire, 
we obtain the conductance for the nanotube systems. 
As a characteristic result, 
conductive processes in the circumferential direction of the nanotube 
are newly added, 
reflecting the shape of contact points 
between the FM and the nanotube edges.

In the parameter set, the nanotube length is of 10 zigzag lines, 
the transfer integrals in the FM and the nanotube parts are $t$~\cite{tranf}, 
those between the edge carbon atoms and 
their nearest FM atoms are $v=0.1t$, 
the on-site energy of carbon is 0, 
and the on-site energy of the FM and $E_{\mbox{\tiny F}}$ are 
the respective ones used in the case of the atomic wire. 
This carbon nanotube appears to behave nearly as a semiconductor, 
in which transmission between the two edges of the nanotube 
is mainly through a small energy gap 
in the vicinity of $E_{\mbox{\tiny F}}$. 
Originally, however, the nanotube is metallic, 
because it is regarded merely as a zigzag ribbon~\cite{fuji} 
with short periodicity. 
Since wave functions on $E_{\mbox{\tiny F}}$ 
localize at the left or right edges of the nanotube~\cite{fuji}, 
they contribute little to the conduction between the two edges.

In the upper panel of Fig. 3, 
we show the MR ratio vs $\frac{N_{\mbox{\tiny path}}}{N_x}$ 
for the cases of $N_{\mbox{\tiny cir}}$=5 ($r$=1), 
10 ($r$=2), and 15 ($r$=3). 
The MR ratios qualitatively exhibit the same dependence on 
$\frac{N_{\mbox{\tiny path}}}{N_x}$ 
as that in the case of atomic wires. 
We emphasize that 
the MR ratio is larger than that for atomic wires, 
and furthermore, 
the MR ratio at each $N_{\mbox{\tiny path}}$ becomes large 
as $N_{\mbox{\tiny cir}}$ increases. 
For $N_{\mbox{\tiny path}} > 1$, 
the above properties mainly are a result of 
the contribution of the CC term of the P case, 
where the magnitude of the CC term based on the IC term 
becomes obviously large 
with increasing $N_{\mbox{\tiny cir}}$ (see lower panel of Fig. 3). 
In particular, we compare 
the nanotube case with the wire case. 
At the points of contact with the FM, 
the nanotubes have a distribution of atoms also in the $y$ direction, 
which is not present in the wires. 
Then, a new conductive process with consistency of $k_y$ 
between the left and right FMs becomes predominant, 
indicating that CC is favored in the $y$ direction 
as well as the $x$ direction. 
In a classical picture, 
when an electron with a certain momentum goes into the nanotubes, 
the momentum in the $y$ direction also tends to be conserved, 
because conduction along the circumference is possible. 
Of course, 
such a conduction in the circumferential direction 
enhances the MR ratio in the $N_{\mbox{\tiny cir}}$=1 case, too.

Finally, we give several comments. 
First, in this study, 
the spin polarization of the FM electrode 
comes from only the single orbitals of the FM atoms, 
although the spin polarization rate~\cite{jul} 
is actually influenced so strongly by interfacial states 
that even its sign can be altered~\cite{Zhao2}. 
We note, however, that 
if the interface of the FM is uniform, 
an increase of the magnitude of the MR ratio 
due to the interference effect can occur. 
Second, the present phenomena are found for numerous similar systems. 
For example, 
MR properties calculated for FM/BN nanotubes/FM junctions 
with insulating nanotubes are almost the same as those 
for the junctions with carbon nanotubes~\cite{kokado}. 
We speculate that 
when many conductive paths are randomly located in the $xy$-plane~\cite{many}, 
larger MR ratios than the present ones can be expected, 
because the CC will appear as predominant features 
in the $y$ direction as well as in the $x$ direction. 
Third, 
we mention the Luttinger liquid (LL) behaviors~\cite{tl}, 
which are characteristic of one-dimensional conductive systems 
with electron-electron interactions. 
Actually, the LL behaviors may exert little influence on the MR ratio and 
its $N_{\mbox{\tiny path}}$ dependence, 
because 
the present MR effect is mainly realized 
by the difference in the spin-dependent potential between the two FMs, 
and also, it does not change without the introduction of spin-polarized paths. 
The LL behaviors may, rather, be observed 
in the temperature dependence on the conductance 
for each magnetization configuration, for example.

In conclusion, 
the spin-dependent transport in the spin-polarized junctions 
of ferromagnetically contacting multiple conductive paths 
was theoretically investigated 
to elucidate the intrinsic relation between 
the number of paths and the conduction, 
and to enhance the MR ratio. 
When many paths are randomly located between the two FMs, 
CC appears as a result of 
electronic wave interference between the two FMs. 
With increasing $N_{\mbox{\tiny path}}$, 
CC becomes more marked, and the MR ratio increases. 
Furthermore, the carbon nanotube junctions show larger MR ratio 
compared with the atomic wire junctions, 
because the intralayer momentum tends to be well conserved 
owing to conduction along the circumference of the nanotubes. 
We expect that the characteristic phenomena 
will be observed with progress of experimental techniques 
and will be utilized 
in nanowire or nanotube MR devices in the future.

This work has been supported by Special Coordination Funds 
for Promoting Science and Technology, Japan.

\newpage
\begin{figure}[ht]
\begin{center}
\epsfxsize=20cm \epsfbox{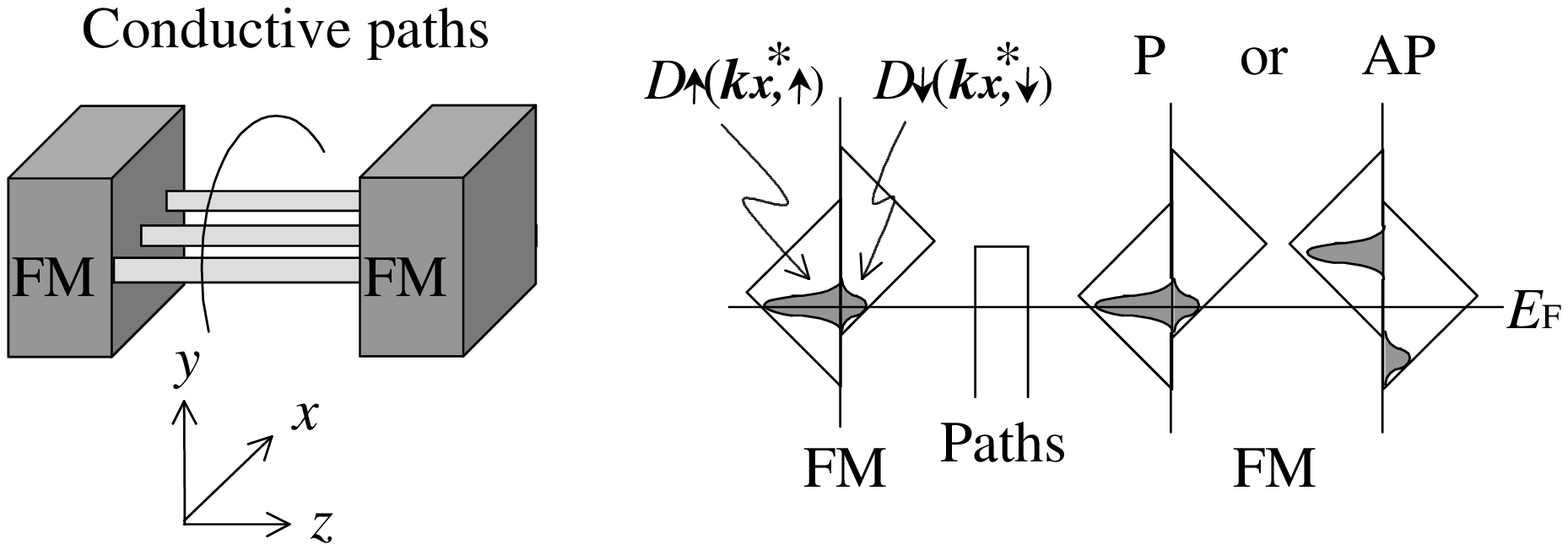} \\[-7.5cm]
\caption{
Left panel: 
A schematic illustration of FM/many paths/FM junctions. 
Right panel: 
A schematic illustration of the local DOS at the interfacial layer 
of the left and right FMs 
for the P or AP magnetization configuration. 
In discussing IC, 
we focus on the structure of the DOS at $E_{\mbox{\tiny F}}$.  
In contrast, for CC, 
we pay attention to the DOS indexed by $k_{x,\uparrow (\downarrow)}^*$, 
$D_{\uparrow (\downarrow)}(k_{x,\uparrow (\downarrow)}^*)$, 
where $k_{x,\uparrow (\downarrow)}^*$ is the FM wave vector 
in the $x$ direction for up (down) spin giving energy levels
in the vicinity of $E_{\mbox{\tiny F}}$ for the P configuration, 
and its dominant component is shown by the shaded region. 
}
\end{center}
\end{figure} 

\newpage

\begin{figure}[ht]
\begin{center}
\vspace*{-10cm}
\epsfxsize=15cm \epsfbox{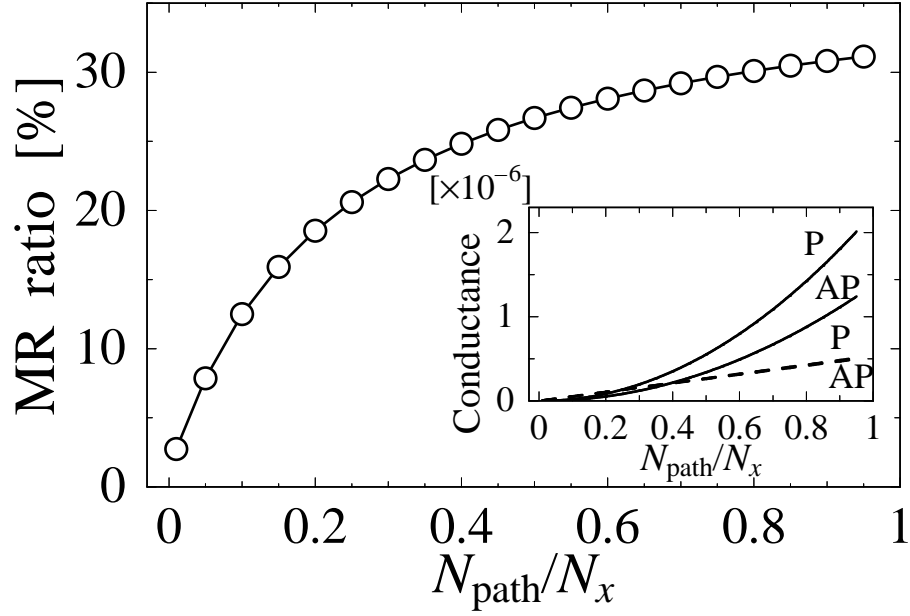} \\
\caption{
MR ratio vs $N_{\mbox{\tiny path}}/N_x$ 
for FM/atomic wires/FM junctions. 
$N_{\mbox{\tiny path}}$=1 corresponds to 
$N_{\mbox{\tiny path}}/N_x$=0.01. 
Inset: The CC and IC terms in $\sum_\sigma \Gamma_\sigma/N_x$ 
vs $N_{\mbox{\tiny path}}/N_x$. 
The solid (dotted) line represents the CC (IC) term. 
}
\end{center}
\end{figure}

\newpage

\begin{figure}[ht]
\begin{center}
\vspace*{-10cm}
\hspace*{-2cm}\epsfxsize=15cm \epsfbox{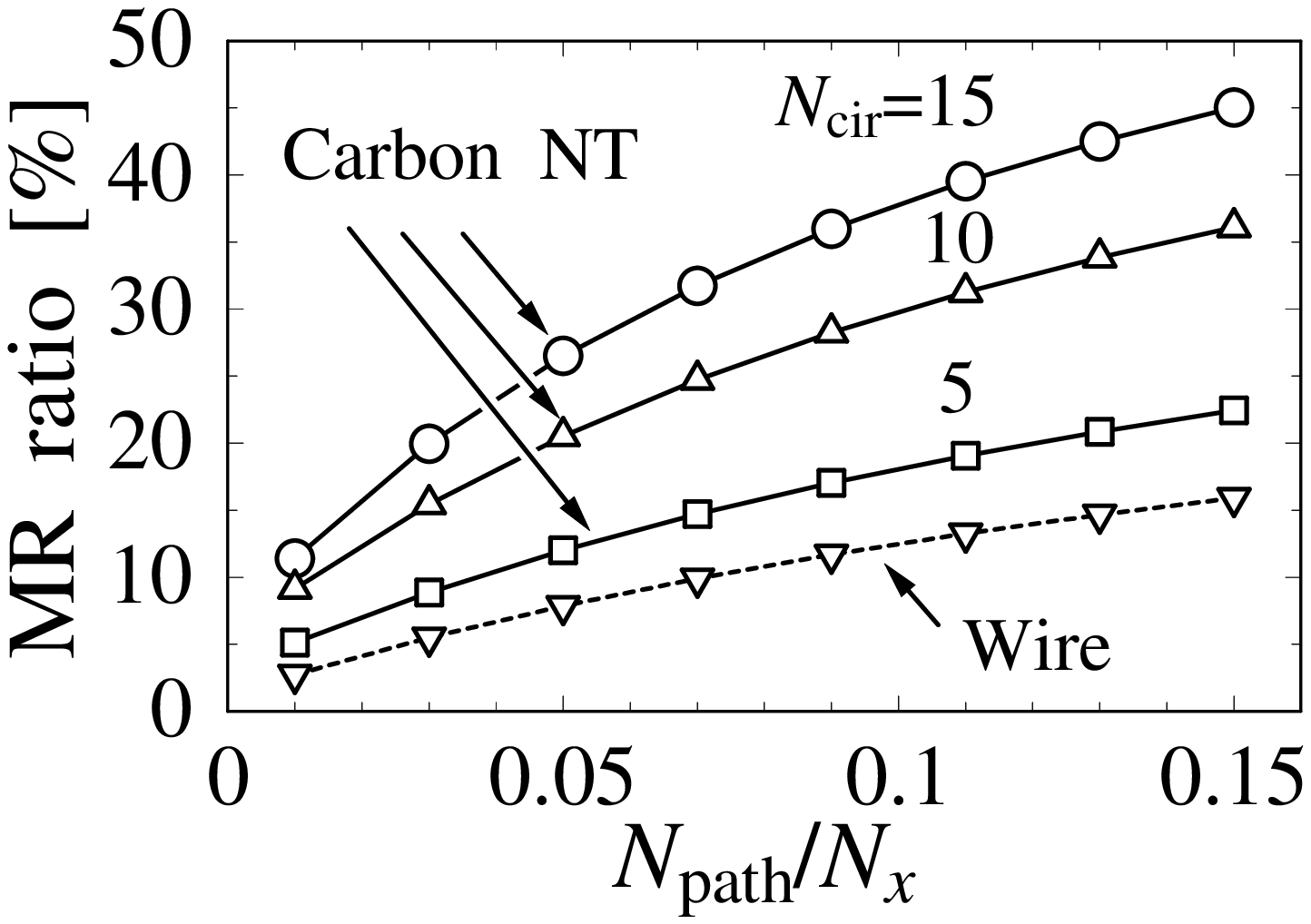} \\[-6cm]
\hspace*{-2cm}\epsfxsize=10cm \epsfbox{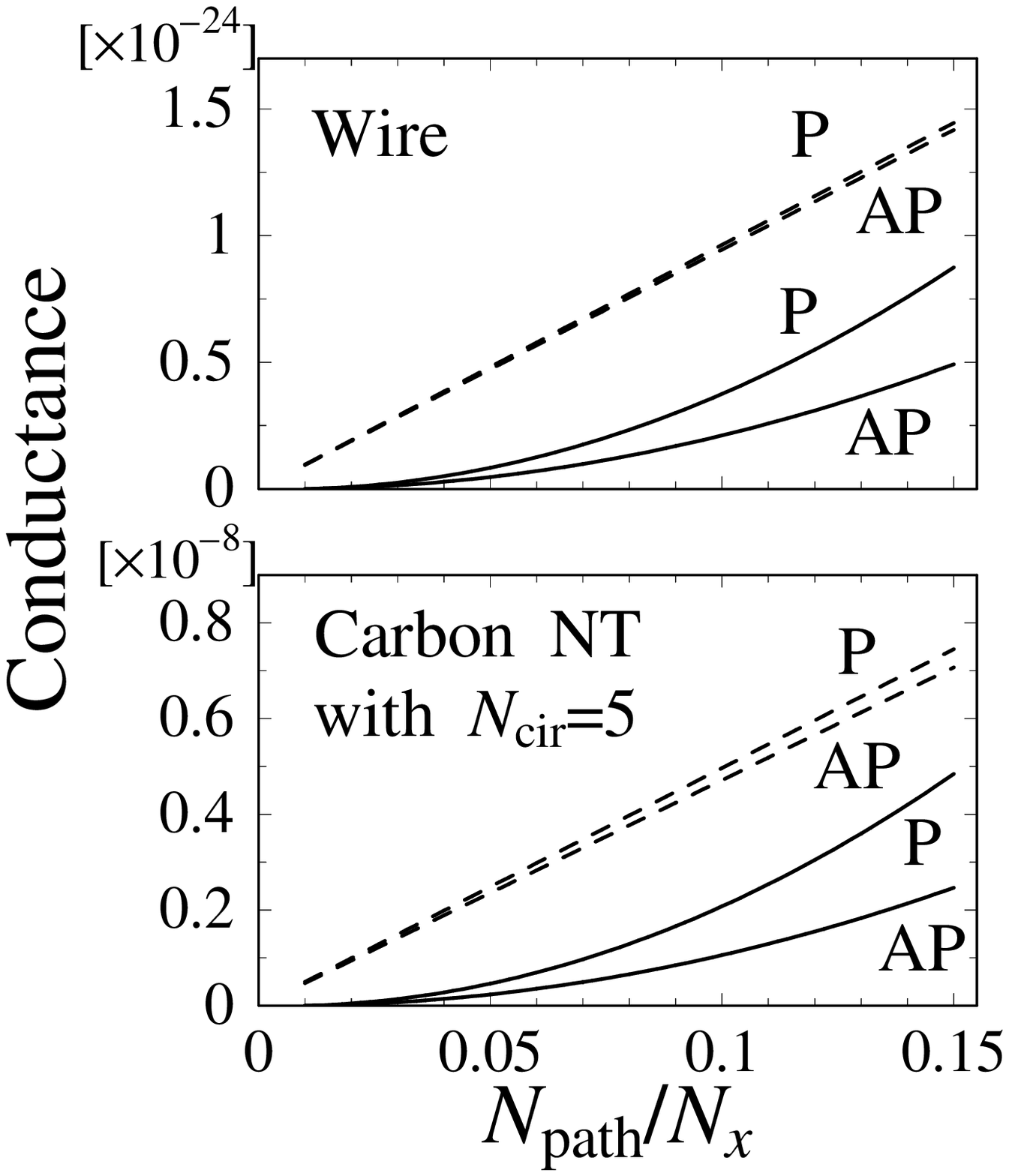} 
\hspace*{-2cm}\epsfxsize=10cm \epsfbox{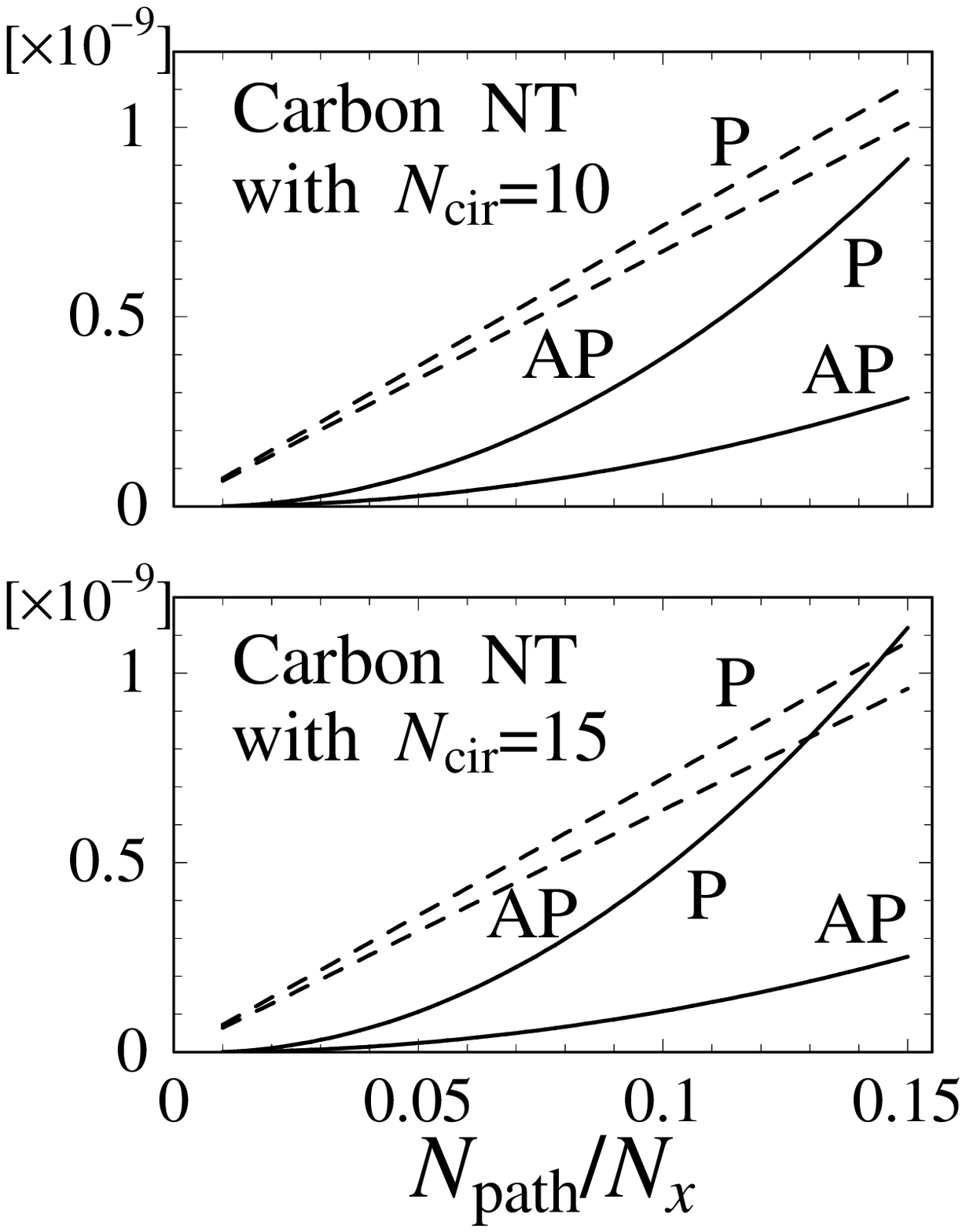} \\
\caption{
Upper panel: MR ratio vs $N_{\mbox{\tiny path}}/N_x$ 
for FM/carbon nanotubes (NT)/FM junctions (solid lines) and 
FM/atomic wires/FM junctions (dotted line). 
$N_{\mbox{\tiny path}}$=1 corresponds to 
$N_{\mbox{\tiny path}}/N_x$=0.01. 
Furthermore, the range of $N_{\mbox{\tiny path}}/N_x$ is determined 
so as to be less than the system size with $N_{\mbox{\tiny cir}}$=15 ($r$=3). 
Lower panel: The CC and IC terms in $\sum_\sigma \Gamma_\sigma/N_x$ 
vs $N_{\mbox{\tiny path}}/N_x$ for the respective cases. 
The solid (dotted) line represents the CC (IC) term. 
}
\end{center}
\end{figure}

\begin{thebibliography}{9}
\bibitem{tranf}
$t$ is now assumed to be $-$2.75 eV~\cite{th1}.
\end{thebibliography}
\end{document}